# Dynamics of interaction of radially symmetric topological solitons in two-dimensional nonlinear sigma model

F. Sh. Shokirov

S. U. Umarov Physical-Technical Institute of Academy of Sciences of the Republic of Tajikistan, Aini ave. 299/1, Dushanbe, 734063, Tajikistan

E-mail: farhod0475@gmail.com

**Abstract:** By methods of numerical simulations the dynamics of interaction of radially symmetric Bellavin-Polyakov type topological vortex in (2+1)-dimensional O(3) nonlinear sigma model is investigated. Obtained numerically the model of topological vortex decay for different values of radius of ring-shaped structure of their energy density onto the localized perturbations, where the sum of Hopf index is preserved. It is shown that the stability of topological solitons, in particularly, depends on the values of radius of ring-shaped structure of their energy density.

## 1. Introduction

This paper presents a numerical investigation of head-on collisions of radially symmetric Bellavin-Polyakov type topological vortices (topological solitons, TS) [1-3] in (2+1)-dimensional anisotropic O(3) nonlinear sigma model (NLSM). The lagrangian and hamiltonian density of two-dimensional O(3) NLSM for the anisotropic case [4-9] in isospin parameterization are written in the next form:

$$\mathcal{L} = \frac{1}{2}\left[\partial_\mu s_a \partial^\mu s_a + (s_3^2 - 1)\right], \tag{1}$$

$$\mathcal{H} = \frac{1}{2}\left[(\partial_0 s_a)^2 + (\partial_k s_a)^2 + (1 - s_3^2)\right],$$

where $\mu = 0,1,2$; $a = 1,2,3$; $s_a s_a = 1$, $k = 1,2$. The Euler-Lagrange equations of the model (1) are

$$2\partial_\mu \partial^\mu \theta + \sin 2\theta \left(1 - \partial_\mu \varphi \partial^\mu \varphi\right) = 0, \tag{2}$$

$$2\cos\theta\, \partial_\mu \varphi \partial^\mu \varphi + \sin\theta\, \partial_\mu \partial^\mu \varphi = 0, \quad \mu = 0,1,\ldots,D, \quad D = 2,$$

where $\theta(x,y,t)$ and $\varphi(x,y,t)$ are the Euler angles [4-9]. Numerical study of interaction of the TS of model (1)-(2) with nonzero Hopf index $Q_t$ (topological charge, TC) [1-3] of the form

$$\theta = 2arctg(r/R)^{Q_t}, \quad \varphi = Q_t\chi - \omega\tau, \tag{3}$$
$$r^2 = x^2 + y^2, \quad \cos\chi = x/r, \quad \sin\chi = y/r,$$

is conducted at $R = \left\{\frac{1}{2}, 1, 2\right\}$, $\omega = 1$ and $Q_t = 3$. A model of two-soliton head-on collision of TS of form (3) of model (2), which moving in the opposite direction at a speed $v(t_0) = (0.0, 0.2)$ at different $\mathcal{R}_{DH}$ values is investigated. We note that our numerical models are based on the three-layer finite difference scheme of the second order of accuracy $O(h^2 + \tau^2)$ [10,11], with using stereographic projection and taking into account the group-theoretic properties of class of O(N) NLSM of the field theory [4-9]. The movement of the TS form of (3) was given by the Lorentz transformation based on the properties of Lorentz invariance of O(3) NLSM (1)-(2). The models of head-on collisions and reflections (with elastic particle-like behavior), and also the decay of the radially symmetric TS of form (3) of model (2) onto the localized perturbations (LP) is obtained.

## 2. Topological solitons-vortices with different $\mathcal{R}_{DH}$ values

In the investigation of numerical models as a trial functions were used the TS of form (3) of model (2) for three values of the radius of the ring-shaped structure of their energy density $(DH)$: $\mathcal{R}_{DH} = r/2$, $\mathcal{R}_{DH} = r$, $\mathcal{R}_{DH} = 2r$. At the $\mathcal{R}_{DH} = r$ ($R_{RL} = 1$) the interaction dynamics of the TS of form (3) have been investigated in our previous works (see, e.g. [4-9]). There were found a number of properties, including: the decay of the TS onto the localized perturbations (LP) with preserving value of $Q_t$; phased annihilation of the TS by the units of $Q_t$ (by the radiation of energy in the form of linear perturbation waves, propagating at the speed of $c$); long-range TS; mutual attraction and repulsion of TS, etc. The energy integral $(En)$ of the obtained models in works [4-9] of interacting TS has been conserved with good accuracy: $\frac{\Delta En}{En} \approx 10^{-6} - 10^{-3}$.

Fig.1 shows the evolution of (energy density $(DH)$ and its contour projection) of a frontal collision of TS of form (3) in (2+1)-dimensional O(3) NLSM (1)-(2), at $\mathcal{R}_{DH} = r/2$ (Fig.1a), $\mathcal{R}_{DH} = r$ (Fig.1b) and $\mathcal{R}_{DH} = 2r$ (Fig.1c). Increasing values of $\mathcal{R}_{DH}$ leads to an increase in TS energy $(En)$ and density $(DH)$ of their energy (but to decrease of the maximum values of DH) assembled in a ring-shape structure (Fig.1).

Next, we present the results of numerical investigation of interaction (head-on collisions) TS of form (3) of model (2) (shown in Fig.1) in the following views:
- $\mathcal{R}_{DH} = r/2 \rightarrow\leftarrow \mathcal{R}_{DH} = r/2$;
- $\mathcal{R}_{DH} = 2r \rightarrow\leftarrow \mathcal{R}_{DH} = 2r$;
- $\mathcal{R}_{DH} = r/2 \rightarrow\leftarrow \mathcal{R}_{DH} = r$;
- $\mathcal{R}_{DH} = r \rightarrow\leftarrow \mathcal{R}_{DH} = 2r$;

at speeds of moves TS are within $v_{RL}(t_0) \in (0.0, 0.2)$.

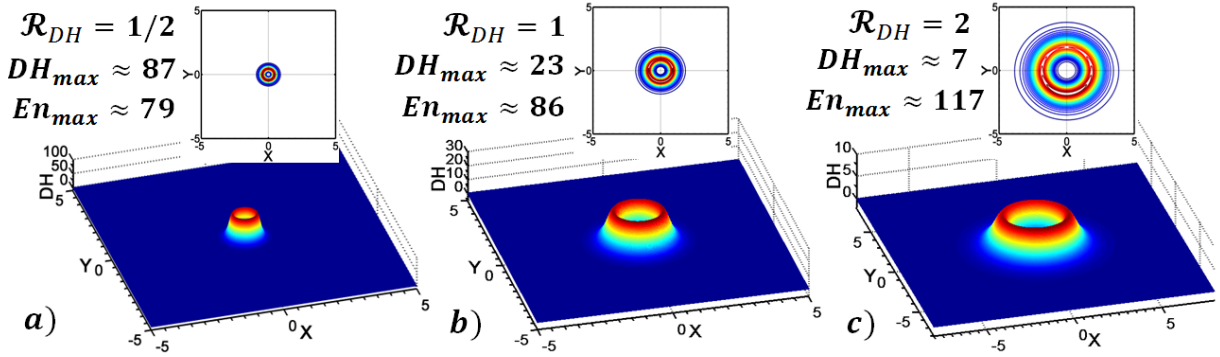

**Fig.1.** Numerical model of evolution ($DH$ and its contour view) of TS of form (3) with the Hopf index $Q_t = 3$, at $t = 0.0$: *a)* $\mathcal{R}_{DH} = r/2$; *b)* $\mathcal{R}_{DH} = r$; *c)* $\mathcal{R}_{DH} = 2r$.

### 3. Interaction of two-dimensional radially symmetric topological solitons

Below present the results of research of models of a head-on collisions of TS of form (3) in anisotropic (2+1)-dimensional O(3) NLSM with different $\mathcal{R}_{DH}$. values of the radius of the ring-shaped structure of their $DH$.

**Case:** $\mathcal{R}_{DH} = r/2 \rightarrow\leftarrow \mathcal{R}_{DH} = r/2$ ($R = 2$). Fig.2 shows the evolution of (energy density ($DH$) and its contour projection) of a frontal collision of TS (3) in (2+1)-dimensional O(3) VNSM, having a value of $\mathcal{R}_{DH} = r/2$. In this case, the radius of the annular structure of energy density ($DH$) of TS almost half (Fig.1a) with respect to the case of $\mathcal{R}_{DH} = r$ (Fig.1b), the dynamics of interaction, which was studied in our previous works [4-9]. So, the TS moving in the opposite direction (Fig.2), where the initial velocity ($t_0 = 0.0$), given by the Lorentz transformation for both TS is equal: $|v_{RL}(t_0)| \approx 0.0995$. Simulation time: $t \in [0.0, 90.0]$. Until the moment of collision (at $t = 42.3$, Fig.2b) are both TS pass the distance equal to $s \approx 4.0$ units almost with speed given at $t_0 = 0.0$. At the collision in the interval $t \in (42.3, 55.0)$ the interacting TS form a bound state (see, Fig.2c at $t = 50.1$), which after this time are decay onto the 4 LP (see, Fig.2d at $t = 60.0$). The formed LP has the topological charge (Hopf index) equal to a $Q_t = 1$ (two LP) and $Q_t = 2$ (two LP), i.e., the sum of TC of system of interacting TS is preserved. These LP the formed at $t \approx 45.0$, remains stable until the time $t = 90.0$ (Fig.2e) and continue to move to the opposite sides from the resonance zone. The trajectory of move of formed LP due to the chiral field of TS changed symmetrically at a specific angle $\alpha$ (Fig.2e). The energy integral system of interacting TS before the collision process (at $t \approx 45.0$) is maintained with high accuracy $\frac{\Delta En}{En} \approx 10^{-6}$ (Fig.2f). Recall that the edges of the area of simulation are set the special boundary conditions, which absorb the excess energy emitted by interacting TS as form of linear perturbation waves. We note that in the case of $\mathcal{R}_{DH} = r$ at the similar speeds of TS the decay of solitons onto the LP was not observed – TS after the collision are reflected from each other and continued to move in opposite directions from the resonance zone [4-9].

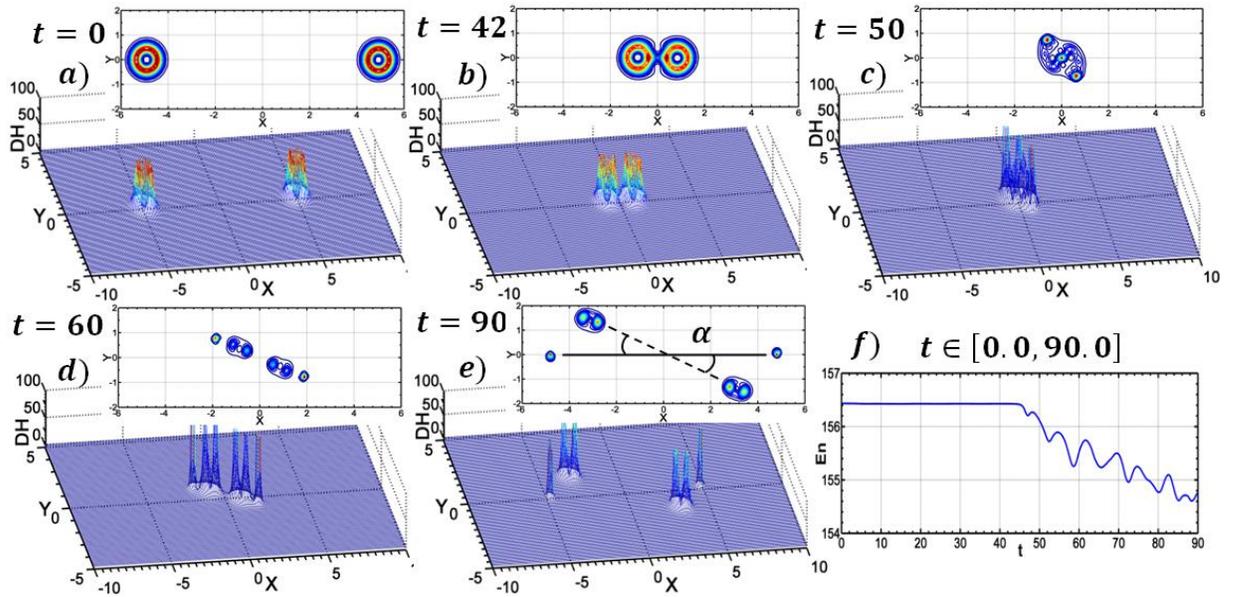

**Fig.2.** Numerical model of evolution ($DH$ and its contour view) of head-on collision of TS of form (3) with the Hopf index $Q_t = 3$ at $|\vec{v}_{LR}(t_0)| \approx 0.0995$ and $\mathcal{R}_{DH} = r/2$: *a)* $t = 0.0$; *b)* $t = 42.3$; *c)* $t = 50.1$; *d)* $t = 60.0$; *e)* $t = 90.0$; *f)* integral of energy ($En$): $t \in [0.0, 90.0]$.

The increase in the speed of TS of form (3) the process of the interaction of which is shown in Fig.2 did not give a qualitatively new result, solitons at the collision are decay onto the LP, where also preserved the sum of $Q_t$.

**Case: $\mathcal{R}_{DH} = 2r \rightarrow\leftarrow \mathcal{R}_{DH} = 2r$ ($R = 1/2$).** In this case, at the speed of TS equal to $|v_{RL}(t_0)| \approx 0.0995$ the type of interaction is different from that of the process shown in Fig.2 of the previous experiment. In the case $\mathcal{R}_{DH} = 2r$ radius of the ring-shaped structure of the energy density ($DH$) of TS almost twice as much (Fig.1c) relative to the case of $\mathcal{R}_{DH} = r$ (Fig.1b), the dynamics of interaction, which is studied in our previous works [4-9]. The collision of TS in this case ($\mathcal{R}_{DH} = 2r$) is similar to the case of $\mathcal{R}_{DH} = r$ [4-9] – after collisions TS are reflected and preserving stability, continue to move in the opposite sides from the resonance zone, along a deflected trajectory on a certain angle.

Next, consider the numerical model similar to the previous process of head-on collision, but with twice the speed of moves interacting TS: $|v_{RL}(t_0)| \approx 0.196$. At these speeds, in the case $\mathcal{R}_{DH} = r$ we were observed the decay of TS onto the LP [4-9]. But in this case ($\mathcal{R}_{DH} = 2r$) TS showing the stability and the process of interaction occurs without decay of the TS, similar to the processes described above when $|v_{RL}(t_0)| \approx 0.0995$ – solitons a preserved the integrity, are reflected each other. Below we give a more detailed description of this process (Fig.3).

Fig.3 shows the initial state ($t_0 = 0.0$) of numerical model of interaction of TS when $\mathcal{R}_{DH} = 2r$ (the evolution of the energy density ($DH$) and its contour projections). Simulation time: $t \in [0.0, 72.0]$; the speed of are both TS is equal to $|v_{RL}(t_0)| \approx 0.196$. Until the moment of collision (Fig.3b) at the $t = 38.1$, are both TS pass the distance equal to about $s \approx 5.1$ units, the loss of the average speed of TS is significant and equal $v_{loss}(t = 38.1) \approx 31.7\%$.

At the $t = 46.2$, are both interacting TS to form a bound state, in the center which is concentrated of a dense concentration of energy (Fig.3c). Next, when $t > 55.0$ occurs the separated (reflection) of TS from each other (Fig.3d) and TS, while preserving stability, continue to move in the opposite sides from the resonance zone, along a deflected trajectory on a certain angle (Fig.3e). The energy integral of system of interacting TS is preserved with accuracy $\frac{\Delta En}{En} \approx 10^{-3}$ (Fig.3f).

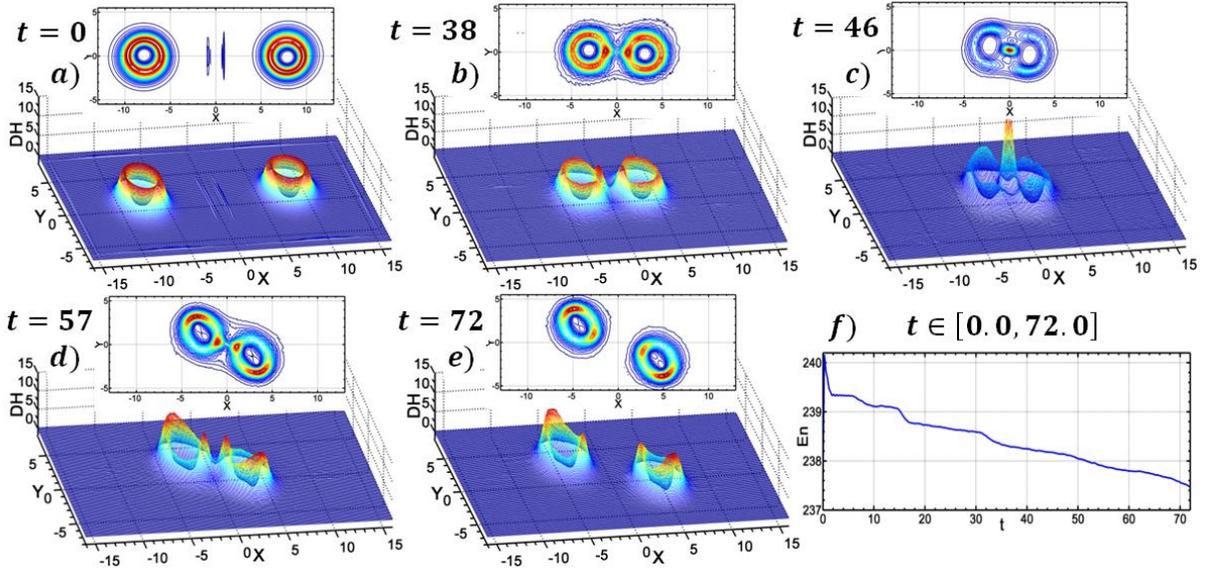

**Fig.3.** Numerical model of evolution ($DH$ and its contour view) of head-on collision of TS of form (3) with the Hopf index $Q_t = 3$ at $|\vec{v}_{LR}(t_0)| \approx 0.196)$ and $\mathcal{R}_{DH} = 2r$: *a)* $t = 0.0$; *b)* $t = 38.1$; *c)* $t = 46.2$; *d)* $t = 57.9$; *e)* $t = 72.0$; *f)* integral of energy ($En$): $t \in [0.0, 72.0]$.

The conducted experiments show that TS of form (3) of model (2) in the case of $\mathcal{R}_{DH} = 2r$ are more stable with respect to cases $\mathcal{R}_{DH} = r$ and $\mathcal{R}_{DH} = r/2$, i.e., increasing the radius of the ring-shape structure of the energy density ($DH$) leads to a corresponding increase of TS stability.

**Case:** $\mathcal{R}_{DH} = r/2 \rightarrow\leftarrow \mathcal{R}_{DH} = r$ ($\mathcal{R}_{RIGHT} = 2$, $\mathcal{R}_{LEFT} = 1$). In this case, at the speed of TS equal to $|v_{RL}(t_0)| \approx 0.0995$ the interaction process are similar to the process described in Fig.3 of the previous experiment. After the collision TS are reflected and preserving stability, continue to move in the opposite sides of the resonance zone, along the rejected trajectory by a certain angle.

Now consider the same model of a frontal collision with a twice speed of interacting TS: $|v_{RL}(t_0)| \approx 0.196$. Fig.4 shows the moving TS in opposite directions, where the initial (Fig.4a) speed ($t_0 = 0.0$), given by the Lorentz transformation for both TS is: $|v_{RL}(t_0)| \approx 0.196$. The topological charge (Hopf index) interacting TS is $Q_t = 3$, simulation time $t \in [0.0, 52.2]$. Until the moment of collision (Fig.4b), at $t = 21.0$, are both TS pass the distance equal to about $s \approx 3.8$ units. After the collision, at the time $t \approx 30.0$ (Fig.4c), there is a process of decay of both TS onto the 6 LP, each of which has a single Hopf index $Q_t = 1$ (the sum of TC $Q_t = 6$ of system of interacting TS are preserved). These LP formed a during the $t \approx 30.0$, remains stable until the time $t \approx 52.2$ (Fig.4de) and continue to move to the sides from the resonance zone. The trajectories of the formed LP as a result of the chiral field of TS are rejected at different angles (Fig.4de) and are not symmetrical about the center of the collision area. The latter circumstance is explained by the different values of $\mathcal{R}_{DH}$ interacting TS. The energy integral system of interacting TS until the collision moment ($t < 21.0$) is preserved with high accuracy $\frac{\Delta En}{En} \approx 10^{-6}$ (Fig.4f).

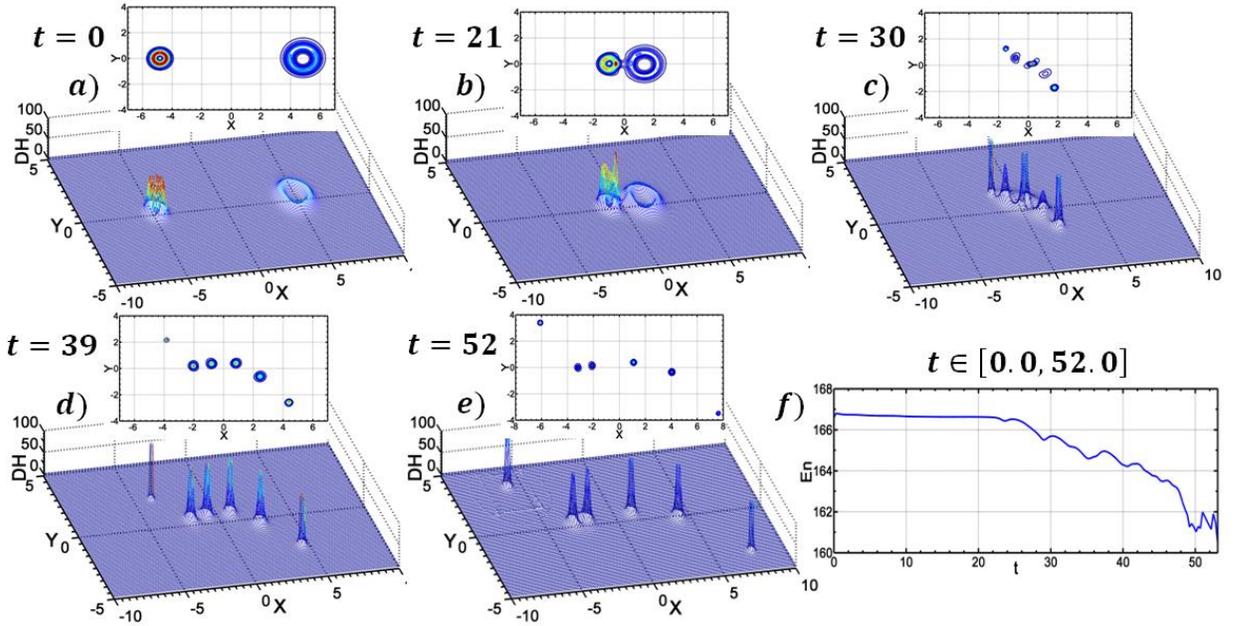

**Fig.4.** Numerical model of evolution ($DH$ and its contour view) of head-on collision of TS of form (3) with the Hopf index $Q_t = 3$ at $|\vec{v}_{LR}(t_0)| \approx 0.196$), $\mathcal{R}_{DH} = r/2$ (left) and $\mathcal{R}_{DH} = r$ (right): *a)* $t = 0.0$; *b)* $t = 21.0$; *c)* $t = 30.0$; *d)* $t = 39.9$; *e)* $t = 52.2$; *f)* integral of energy ($En$): $t \in [0.0, 52.0]$.

**Case:** $\mathcal{R}_{DH} = r \rightarrow\leftarrow \mathcal{R}_{DH} = 2r$ ($\mathcal{R}_{RIGHT} = 1$, $\mathcal{R}_{LEFT} = 1/2$). In these experiments at the speeds of TS equal to $|v_{RL}(t_0)| \in (0.0, 0.197)$ the interaction process are similar to the process described in Fig.3. After collisions TS are reflected and preserving stability, continue to move in the opposite sides from the resonance zone, along a deflected trajectory on a certain angle. The energy integral system of interacting TS is preserved with accuracy: $\frac{\Delta En}{En} \approx 10^{-5} - 10^{-4}$.

## 4. Conclusion

Numerical experiments carried out in this paper show that the process of interaction of TS of form (3) of model (2) at $\mathcal{R}_{DH} = r/2$, $\mathcal{R}_{DH} = r$ and $\mathcal{R}_{DH} = 2r$ are similar to the interaction processes between the TS with $\mathcal{R}_{DH} = r$, which have been studied in detail in previous papers [4-9]. The new property of the TS detected at the studied of their interactions in this paper is a dependence of the TS stability from the $\mathcal{R}_{DH}$ value. So, the TS with a relatively little $\mathcal{R}_{DH}$ value at low speeds of moves, at interactions the decayed onto the LP. (See, for example, Fig.2 and Fig.4). While the TS with a relatively high $\mathcal{R}_{DH}$ value at low speeds of interaction retain integrity – they are reflected upon collision with each other, keep a stable state and move in opposite directions (see, e.g., Fig.3). In the case of the decay of the TS onto the LP (see, for example, Fig.2, Fig.4) preserved the property of the constancy of the sum of Hopf index $Q_t$ (found in [4-9]) system of interacting TS of form (3) of anisotropic (2+1)-dimensional O(3) NLSM.

## Acknowledgments

The author is grateful to Prof. Kh.Kh. Muminov for valuable suggestions made during the discussion of the results.